\documentstyle[aaspp4]{article}

\catcode`\@=11 
\def\@versim#1#2{\vcenter{\offinterlineskip
        \ialign{$\m@th#1\hfil##\hfil$\crcr#2\crcr\sim\crcr } }}
\newcommand{\Ref}{\hangindent=20pt \hangafter=1 \noindent}
\newcommand{\StartRef}{\hyphenpenalty=10000 \raggedright}

\newcommand{\beq}{\begin{equation}}
\newcommand{\eeq}{\end{equation}}
\def\lsim{\mathrel{\mathpalette\@versim<}}
\def\gsim{\mathrel{\mathpalette\@versim>}}

\def\@versim#1#2{\vcenter{\offinterlineskip
        \ialign{$\m@th#1\hfil##\hfil$\crcr#2\crcr\sim\crcr } }}
\catcode`\@=12 

\begin{document}

\title{Accretion Disks around Black Holes with Advection and Optical Depth Transition}

\author{Yulia V. Artemova\altaffilmark{1},
Gennadi S. Bisnovatyi-Kogan\altaffilmark{1,}\altaffilmark{2}}

\affil{\altaffilmark{1}Space Research Institute, Profsoyuznaya 84/32, Moscow
117997, Russia;\\
julia@iki.rssi.ru, gkogan@iki.rssi.ru}

\affil{\altaffilmark{2}Joint Institute Nuclear Research, Dubna, Russia}

\author{Igor V. Igumenshchev\altaffilmark{3}}
\affil{\altaffilmark{3}Laboratory for Laser Energetics, University of Rochester,\\
250 East River Road, Rochester, NY 14623, USA ;
iigu@lle.rochester.edu}

\author{Igor D. Novikov\altaffilmark{4,}\altaffilmark{5}}
\affil{\altaffilmark{4}Niels Bohr Institute, Blegdamsvej 17,
DK-2100, Copenhagen, Denmark; novikov@nbi.dk}

\affil{\altaffilmark{5}Astro-Space
Center of P.N. Lebedev Physical Institute, Profsouznaya 84/32,
Moscow 117997, Russia}

\medskip
\setcounter{footnote}{0}

\begin{abstract}

We consider the effects of advection and radial gradients of
pressure and radial drift velocity on the structure of accretion
disks around black holes with proper description of optically thick/thin
transitions.
We concentrated our efforts on the models
with large accretion rate. Contrary to disk models
neglecting advection, we find that continuous solutions extending
from the outer disk regions to the inner edge exist for all
accretion rates we have considered. We show that the sonic point
moves outward with increasing accretion rate, and that in the
innermost disk region advection acts as a heating process that may
even dominate over dissipative heating. Despite the importance of
advection on it's structure, the disk remains geometrically thin.
Global solutions of advective accretion disks, which describe
continuously the transition between optically thick outer region
and optically thin inner region are constructed and analyzed.

\noindent {\em Subject headings:} Accretion, accretion disks ---
black hole physics --- hydrodynamics
\end{abstract}


\section{Introduction}
The ``standard accretion disk model'' of Shakura (1972) and
Shakura \& Sunyaev (1973), that has been widely used to model
accretion flows around black holes, is based on a number of
simplifying assumptions. In particular, the flow is assumed to be
geometrically thin and with a Keplerian angular velocity
distribution. This assumption allows gradient terms in the
differential equations describing the flow to be neglected,
reducing them to a set of algebraic equations, and thereby fixes
the angular momentum distribution of the flow. For low accretion
rates, $\dot M$, this assumption is generally considered to be
reasonable.
   Since the end of the seventies, however, it has been realized that
for high accretion rates, advection of energy with the flow can
crucially modify the properties of the innermost parts of
accretion disks around black holes. A deviation from a Keplerian
rotation may result.

Initial attempts to solve the more general disk problem only
included advection of energy and the radial gradient of pressure
in models with small values of the viscosity parameter,
$\alpha=10^{-3}$ (Paczy\'nski \& Bisnovatyi-Kogan 1981), and it
was shown that including radial velocity in the radial momentum
equation would not change principally the results for such a small
$\alpha$ (Muchotrzeb \& Paczy\'nski 1982). Liang \& Thomson (1980)
emphasised the importance of the transonic nature of the radial
drift velocity, and the influence of viscosity on the transonic
accretion disk solutions was noted by Muchotrzeb (1983), who
claimed that such solutions only existed for viscosity parameters
smaller than $\alpha_{*}\simeq 0.02-0.05$. Matsumoto et al.\
(1984), then showed that solutions with $\alpha>\alpha_{*}$ do in
fact exist, but the nature of the singular point, where the radial
velocity equals the sound velocity, is changing from a saddle to a
nodal type and the position of this point is shifted substantially
outwards in the disk. Matsumoto et al.\ (1984) also demonstrated
the non-uniqueness of the solutions with a nodal type critical
point for given Keplerian boundary conditions at the outer
boundary of the disk (see also Muchotrzeb-Czerny 1986). Extensive
investigation of accretion disk models with advection for a wide
range of the disk parameters, $\dot{M}$ and $\alpha$, was
conducted by Abramowicz et al.\ (1988), with special emphasis on
low $\alpha$. Misra \& Melia (1996) considered optically thin
two-temperature disk models and treated advection in the framework
of the Keplerian disk model, but fixed the proton temperature
somewhat arbitrarily at the outer boundary. Chakrabarti (1996)
solved the advection problem containing shock waves near the
innermost disk region, considering accretion through saddle
points. Numerical solutions of accretion disks with advection have
been obtained by Chen and Taam (1993) for the optically thick case
with $\alpha=0.1$, and for the optically
thin case (Narayan 1996). A simplified account of
advection has recently been attempted, either treating it like an
additional algebraic term assuming a constant radial gradient of
entropy (Abramowicz et al.\ 1995; Chen et al.\ 1995; Chen 1995),
or using the condition of self-similarity (Narayan \& Yi 1994).

Over the last few years it has become clear, that neglecting the
advective heat transport at high $\dot{M}$ leads to qualitatively
wrong conclusions about the topology of the family of solutions of
the disk structure equations (see for example Abramowicz et al.\
1995; Chen et al.\ 1995; Artemova et al.\ 1996). The disk
structure equations without advection give rise two branches of
solutions: optically thick and optically thin, which do not
intersect if $\dot M<\dot M_{cr} \approx (0.6-0.9)  \dot{M}_{\rm
Edd}$ for $\alpha=1$ and $M_{BH}=10^{8}M_{\odot}$, where
$\dot{M}_{\rm Edd}$ is the Eddington accretion rate (Artemova et
al.\ 1996). For larger accretion rates there are no solutions of
these equations extending continuously from large to small radii,
and with Keplerian boundary conditions at the outer boundary of
the disk (see also Liang \& Wandel 1991; Wandel \& Liang 1991; Luo
\& Liang 1994). It was argued by Artemova et al.\ (1996), that for
accretion rates larger than $\dot{M}_{cr}$ advection becomes
critically important and would allow solutions extending all the
way to the inner disk edge also to exist for
$\dot{M}>\dot{M}_{cr}$.

 The consistent transonic solutions for optically thick  advective accretion
disks were constructed in a wide range of values of the viscosity
parameter $\alpha$ and mass accretion rate (see Artemova et al.\
(2001)) and character of singular points of obtained solutions was
investigated.
 However for some values of mass accretion rate and viscosity parameter
the corresponding value of optical depth in the inner region of
accretion disk becomes less than unity and the used
assumptions for optically thick disk approximation are violated. It requires
to include into system of equations the transition formulae for
radiative pressure and radiative flux (Artemova et al.,
1996) to describe correctly the intermediate region  between
optically thick and optically thin zone.

The goal of the present paper is to construct explicitly accretion
disk models with optical depth transition for high $\dot{M}$
taking advective heat transport self-consistently into account. We
also include radial gradients of pressure  and radial drift
velocity and we allow for the non-Keplerian character of the
circular velocity. Furthermore, we use the geometrically thin disk
approximation because, the relative thickness of the disk in the
considering models is less than unity. We show that solutions
extending from large radii to the inner edge of the disk can be
constructed even for accretion rates considerably larger than
$\dot M_{cr}$ and the smooth transition between optically thick and
optically thin regions in accretion disk appeares when the value
of mass accretion rate becomes close to the critical one for
considering models. We find that advection is very important in
the innermost disk region.

In \S 2 we introduce our model and solution methods, while in \S 3
we discuss our results.
\section{The Model and the Method of Solution}

   In this paper we will consider the full set of solutions to
the disk structure equations with advection including the
optically thick, optically thin and the intermediate zones in
accretion disks.

   In our models we used two differents prescriptions for viscosity
$$
t_{r\phi}=-\alpha P ~,   \eqno(1)
$$
suggested by Shakura (1972) and the angular velocity
gradient-dependent viscous stress,
$$
t_{r\phi}=\rho\nu r {d\Omega\over dr} ~,  \eqno(2)
$$
where $\nu$ is the kinematic viscosity coefficient.


 We use from now on geometric units with $G=1$, $c=1$, use
$r$ as the radial coordinate scaled to $r_g=M$, and scale all
velocities to $c$. We work with the pseudo-Newtonian potential
proposed by Paczy\'nski and Wiita (1980), $\Phi=-M/(r-2)$, that
provides an accurate, yet simple approximation to the
Schwarzschild geometry. We normalize the accretion rate as
$\dot{m}=\dot M/\dot M_{\rm Edd}$, where $\dot M_{\rm Edd}=L_{\rm
Edd}=4\pi M m_p/\sigma_T$, in our units.

 We use the same set of equations, ingredients and boundary
conditions in our models as in the paper of Artemova et al. (2001)), except for
changes required by the formulae for radiative flux and radiation
pressure to describe correctly the intermediate zone in the
accretion disk at high accretion rates.
 We used the same numerical method as  Artemova et al. (2001)
to solve this modified system of algebraic and differential
equations.

  The following equations are therefore modified:

  The vertically averaged energy conservation equation:
$$ Q_{adv}=Q^{+}-Q^{-}, \eqno(3) $$ where $$
Q_{adv}=-{\dot{M}\over 4\pi r}\left[{dE\over dr}+P{d\over dr}
\left({1\over\rho}\right)\right], \eqno(4) $$ $$
Q^{+}=-{\dot{M}\over 4\pi} r \Omega{d\Omega\over dr}
 \left(1-{l_{in}\over l}\right),  \eqno(5)
$$
$$
 Q^{-}={2 a T^4 c \over 3 \kappa \rho
h}\left(1+\frac{4}{3\tau_{0}}+ \frac{2}{3\tau_{*}^2}\right)^{-1} ,
\eqno(6)
$$
 are the advective energy, the viscous dissipation
rate and the cooling rate per unit surface, respectively, $T$ is
the midplane temperature, $\kappa$ is the opacity, $a$ is the
radiation constant and $\tau_{0}$ is the Thompson optical depth,
$\tau_{0}=\kappa \rho h$.
 Here we have introduced the total optical depth to absorption,
$\tau_\alpha\, \ll \, \tau_{0}$,

%
$$
\tau_{\alpha}\simeq 5.2*10^{21}{\rho^{2}T^{1/2}h\over acT^{4}} ~~,       \eqno(7)
$$
and the effective optical depth
  \[
  \tau_{*}=\left(\tau_{0}\tau_{\alpha}\right)^{1/2}   .
  \]
 Where $\rho$ is the density and $h$ is the half-thickness of the
 disk.

   The equation of state for the matter consisted of a gas-radiation
mixture is

$$
P_{\rm tot}=P_{\rm gas}+P_{\rm rad}  ~,  \eqno(8)
$$

 where the gas pressure is given by
  \[
  P_{\rm gas}=\rho{\cal R}T ~,
  \]
where ${\cal R}$ is the gas
constant.

 The expression for the radiation pressure is

$$
P_{\rm rad}={a T^4 \over 3}\left(1+\frac{4}{3\tau_{0}}\right)
\left(1+\frac{4}{3\tau_{0}}\textsl{+}\frac{2}{3\tau_{*}^2}\right)^{-1}    .
$$

  In our calculations we use the dimensionless parameter $\beta$ 
to describe the relation of gas pressure to the total pressure in 
accretion disk

$$
\beta= \frac{P_{\rm gas}}{P_{\rm gas}+P_{\rm rad}} ~~ .
$$
  
The specific energy of the mixture is
$$
\textbf{}\rho E={3\over2}P_{\rm gas}+3P_{\rm rad} ~ .     \eqno(9)
$$
  Our method allows us to construct a self-consistent solution to the system of
equations from very large radii, $r>>100$, and down to the
innermost regions of the disk.
\section{Results and Discussion}
 We will now compare the solutions with advection and without it.
 In the ``standard model'', for accretion
rates $\dot{m}<\dot{m}_{cr}=36$, (for $\alpha$=0.5 and
$M_{BH}=10M_{\odot}$, where $\dot{m}_{cr}$ is the critical
accretion rate) there always exist solutions that extend
continuously from large to small radii. When
$\dot{m}>\dot{m}_{cr}=36$ there are no solutions in a range of
radii around $r \approx 13$, and therefore no continuous solutions
extending from large radii to the innermost disk edge (see
detailed discussion by Artemova et al.\ 1996, where however, the
Newtonian potential was used, resulting in $\dot{m}_{cr}=9.4$ for
$\alpha$=1.0 and $M_{BH}=10^{8}M_{\odot}$).

In Figure~1 and Figure~2 we plot the Thomson scattering depth $\tau_{0}$
and the effective optical depth $\tau_{*}$ as
a function of radius, $r$, for different $\dot{m}$ (for
$\alpha$=0.5 and $M_{BH}=10M_{\odot}$), clearly demonstrating that
the solutions to the complete system of disk structure equations
including advection and radial gradients have quite different
properties at high $\dot{m}$ compared to the solutions of the
standard disk model. For $\dot{m}<\dot{m}_{cr}=36$ (dashed lines),
including the gradient terms gives rather small corrections to the
standard disk model. When $\dot{m}>36$ advection becomes essential
and for $\dot{m}>\dot{m}_{cr}$ it changes the picture
qualitatively. When $\dot{m}>\dot{m}_{cr}$ solutions do exist
extending continuously from large radii to the innermost disk
region (the solid lines) where the solution passes through a
"sonic point". However for small values of the mass accretion rate
$\dot{m}<0.1$ all types of solutions (solutions without advection,
optically thick solutions with advection and solutions with
advection and optical depth transition) are very similar and have
the same structure.

  Figure~3 shows the dependence of $\ell_{in}$ on
accretion rate $\dot{m}$ [left panel], and the locations of the inner singular
points $(r_s)_{in}$ as a function of $\dot{m}$ [right panel]
for $\alpha=0.5$ and $M_{BH}=10M_{\odot}$.
Dushed line corresponds to the advective optically thick solutions,
and solid line - to the advective solutions with optical depth
transition.

  Figure~4 shows the dependence of relation of gas pressure to the 
total pressure in  accretion disk, $\beta_s$, on $\dot{m}$
at the inner singular points for the same models as in Figure~3.
The line styles are the same as in Figure~3. The change of value of
$\beta$ form 1 to 0 corresponds to the change of a state from the
gas pressure to radiative pressure dominated one.
The thin disks with $\beta\simeq 1$ are locally stable,
whereas the parts of the disk in which $\beta\simeq 0$
could be thermally and viscously unstable at $\dot{m}\la 100$
(Pringle, Rees, \& Pacholczyk 1973).

  In Figures~5,~6,~7 we plot solutions with advection
for $\dot m=48 > \dot{m}_{cr}$, $\alpha=0.5$, and $M_{BH}=10M_{\odot}$.
  We compare the solution obtained in optically thick approximation
  (dashed line), and the solution with optical depth
transition (solid line).

  In Figure~5 the the smooth transition between optically thick outer region
of accretion disk and optically thin inner region when the accretion rate
is higher than the critical one is demonstrated (the solid line).

   Figure~6 shows the dependence of the temperature on the raduis for the
same model as Figure~5 and the transition from the outer regions with
relatively low temperature to the inner edge of disk with temperature
close to the $10^{9} K$ is clearly seen (solid line).

   The corresponding dependence of density on the raduis is shown on Figure~7.
When the optical depth transition formula is taken into account (solid line)
the density of the disk decreases considerably in the inner regions of accretion
disk, that corresponds to the optically thin high temperature state of the inner
regions.

  Using analysis discussed by Artemova et al.\
(2001) we have determined a type of
singular points in our numerical solutions.

In the case of viscosity prescription (1)
the solutions have two singular points.
The inner singular points, $(r_s)_{in}$, can be saddles
or nodes depending on values of $\alpha$ and $\dot{m}$.
But in the models discussed above, for $\alpha=0.5$ and
$\dot{m}>1.0$, the inner singular point has always nodal-type.
Note that the change of type from the saddle to the nodal one
does not introduce any features in solutions.
The outer singular points, $(r_s)_{out}$, are always of a saddle-type.
In the case of viscosity prescription (2)
the solutions have only inner singular points
which are always of a saddle-type.

  We have obtained unique solutions for structure of advective
accretion disk with optical depth transition in a wide range of
accretion rates and $\alpha$-parameters.
Both viscosity prescriptions (1) and (2)
have been investigated.
The solutions corresponding to both prescriptions are very close
for $\alpha \la 0.1$, and begin to differ at larger $\alpha$.
Due to the technical problems in calculation of the high viscosity
models in the case of viscosity prescription (2) we consetrated our efforts
on the study of high accretion rate models using prescription (1).

The main difference of the present study from the previous ones is
in using more sophisticated numerical technique which accurately
treats the regularity conditions in the inner singular point of
equations. We have performed an analytical expansion at the
singular point to calculate the derivatives of physical
quantities. These derivatives help us to find the proper integral
curve passing through the singular point. Theis approach allow us to
avoid numerical instabilities and inaccuracies, appearing when
only variables at the singular point, but not its derivatives, are
included into a numerical scheme.

\noindent{\it Acknowledgments.}

Yu.A. and G.B.- K. thank RFFI Grant 02-02-16900 for partial
support of this work.

\bigskip\bigskip
{
\footnotesize
\StartRef
\noindent {\large \bf References} \\

%
%
\Ref Abramowicz, M.A., Czerny, B., Lasota, J.P., \& Szuszkiewicz, E.
1988, ApJ, 332, 646 \\

\Ref Abramowicz , M.A., Chen, X., Kato, S., Lasota, J.-P. \& Regev, 
O. 1995, ApJ, 438, L37 \\

\Ref Artemova, I.V.,  Bisnovatyi-Kogan, G.S., Bjornsson, G.,\&
Novikov, I.D. 1996, ApJ, 456, 119 \\

\Ref Artemova, I.V., Bisnovatyi-Kogan, G.S., Igumenschev, I.V., \&
Novikov, I.D. 2001, ApJ, 549, 1050 \\
\Ref Chakrabarti, S.K. 1996, ApJ, 464, 664 \\


\Ref Chen, X., \& Taam, R.E. 1993, ApJ, 412, 254 \\

\Ref Chen, X. 1995, MNRAS, 275, 641 \\

\Ref Chen, X., Abramowicz, M.A., Lasota, J.-P., Narayan, R. \& Yi, I. 1995, ApJ, 443, L61 \\
%

%
%
%
%
\Ref Liang, E.P., \& Thompson, K.A. 1980, ApJ, 240, 271 \\

\Ref Liang, E.P. \& Wandel, A. 1991, ApJ, 376, 746 \\

\Ref  Luo, C. \& Liang, E.P. 1994, MNRAS, 266, 386 \\

\Ref Matsumoto, R., Kato, S., Fukue, J., \& Okazaki, A.T. 1984, PASJ, 36, 71 \\

\Ref Misra, R. \& Melia, F. 1996, ApJ, 465, 869 \\

\Ref Muchotrzeb, B., \& Paczy\'nski, B. 1982, Acta Astr., 32, 1 \\

\Ref Muchotrzeb, B. 1983, Acta Astr., 33, 79 \\

\Ref Muchotrzeb-Czerny, B. 1986, Acta Astr., 36, 1 \\

\Ref Narayan, R. \& Yi, I. 1994, ApJ, 428, L13 \\

\Ref Narayan, R. 1996, in {\it "Basic Physics of Accretion Discs"}, ed. S. Kato et al., (New York : Gordon and Breach) \\

\Ref Paczy\'nski, B., \& Wiita, P.J. 1980, A\&A, 88, 23 \\

\Ref Paczy\'nski, B., \& Bisnovatyi-Kogan, G.S. 1981, Acta Astr., 31, 283 \\

%
%
\Ref Shakura, N.I. 1972, Astron. Zh., 49, 921 \\

\Ref Shakura, N.I., \& Sunyaev, R.A. 1973, A\&A, 24, 337 \\ 

\Ref Wandel, A. \& Liang, E.P. 1991, ApJ, 380, 84

\newpage
\vskip 5in
\newpage

%
\begin{figure}
\plotone{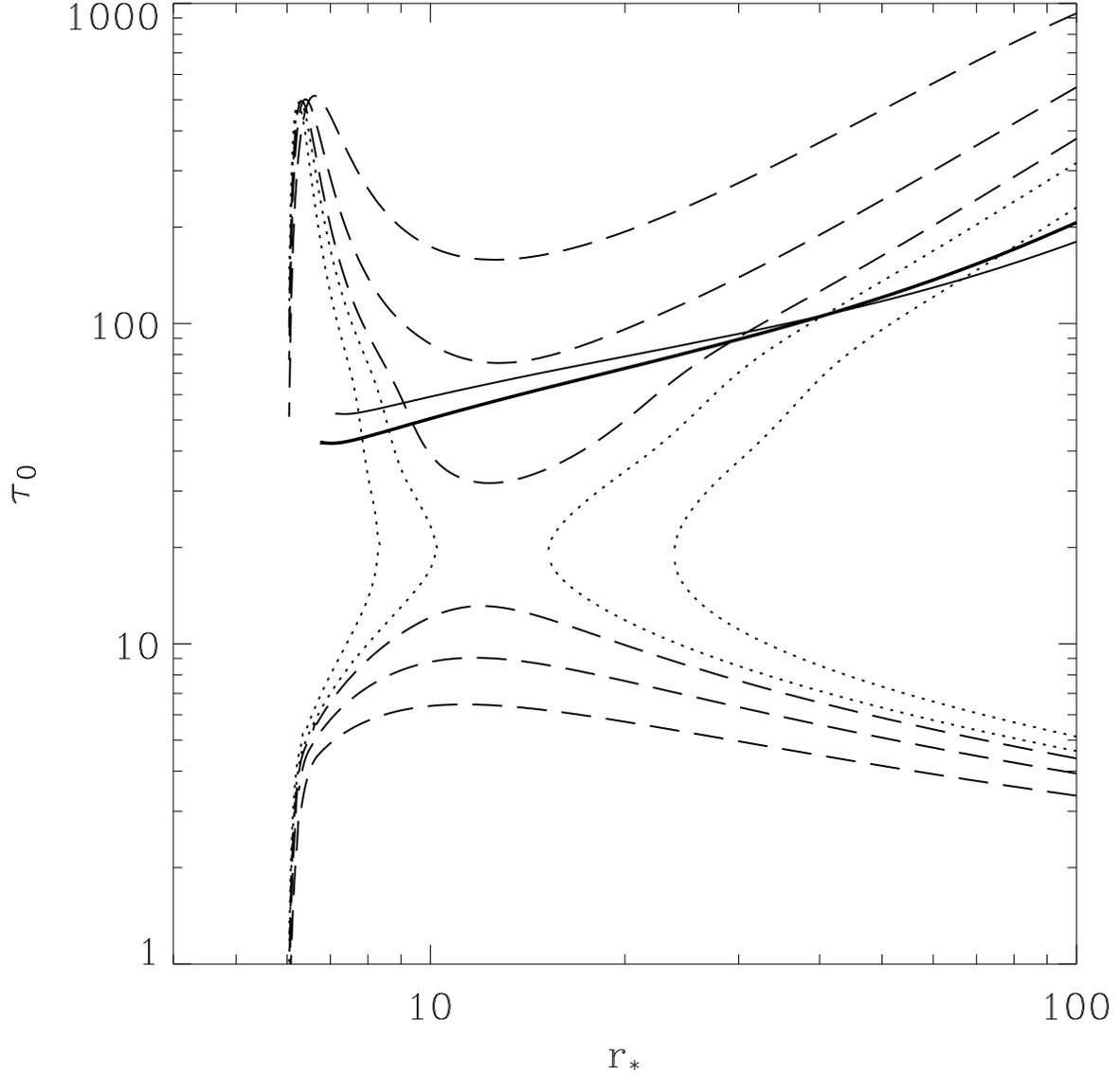}
\caption{The dependence of
the Thomson scattering depth on the radius
for the models with $\alpha=0.5$ and
$M_{BH}=10M_{\odot}$.
Dashed lines correspond to the solutions without advection and
$\dot{m}<\dot{m}_{cr}=36$.
Dotted lines correspond to the non-physical solutions without
advection for $\dot{m}=\dot{m}_{cr}=36$ and $\dot{m}=50$ (from the
center to the edge of the picture respectively). Solid lines
correspond to the solutions with advection and the mass accretion
rate higher than the critical one. Thick solid line corresponds to
$\dot{m}=36.0$, and the thin solid line to $\dot{m}=50.0$.
\label{fig1}}
\end{figure}
\begin{figure}
\plotone{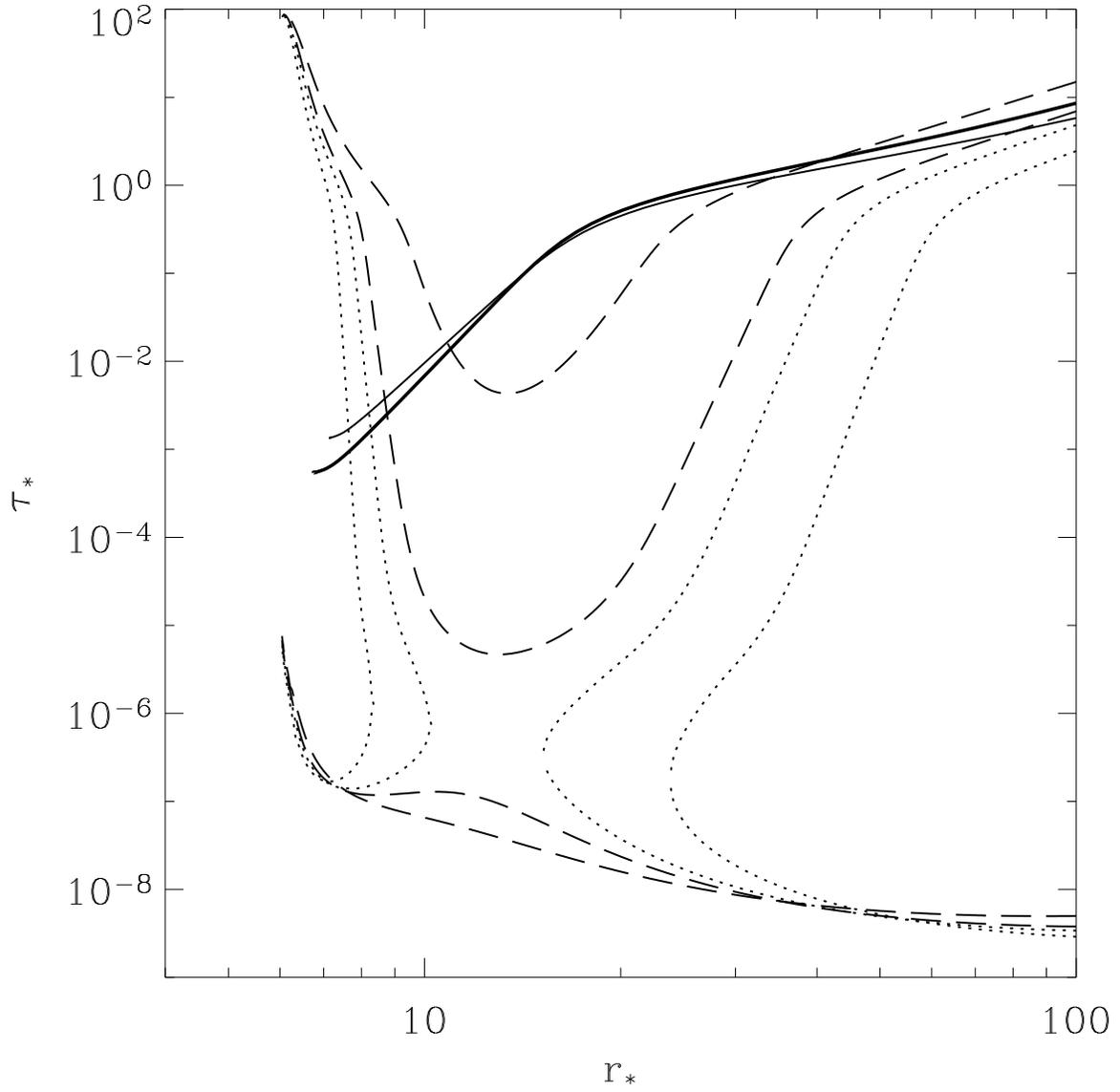}
\caption{The dependence of the effective
optical depth on the radius for the same models as in Fig.1.
The styles of the curves are the same as in Fig.1.
\label{fig2}}
\end{figure}

\begin{figure}
\plottwo{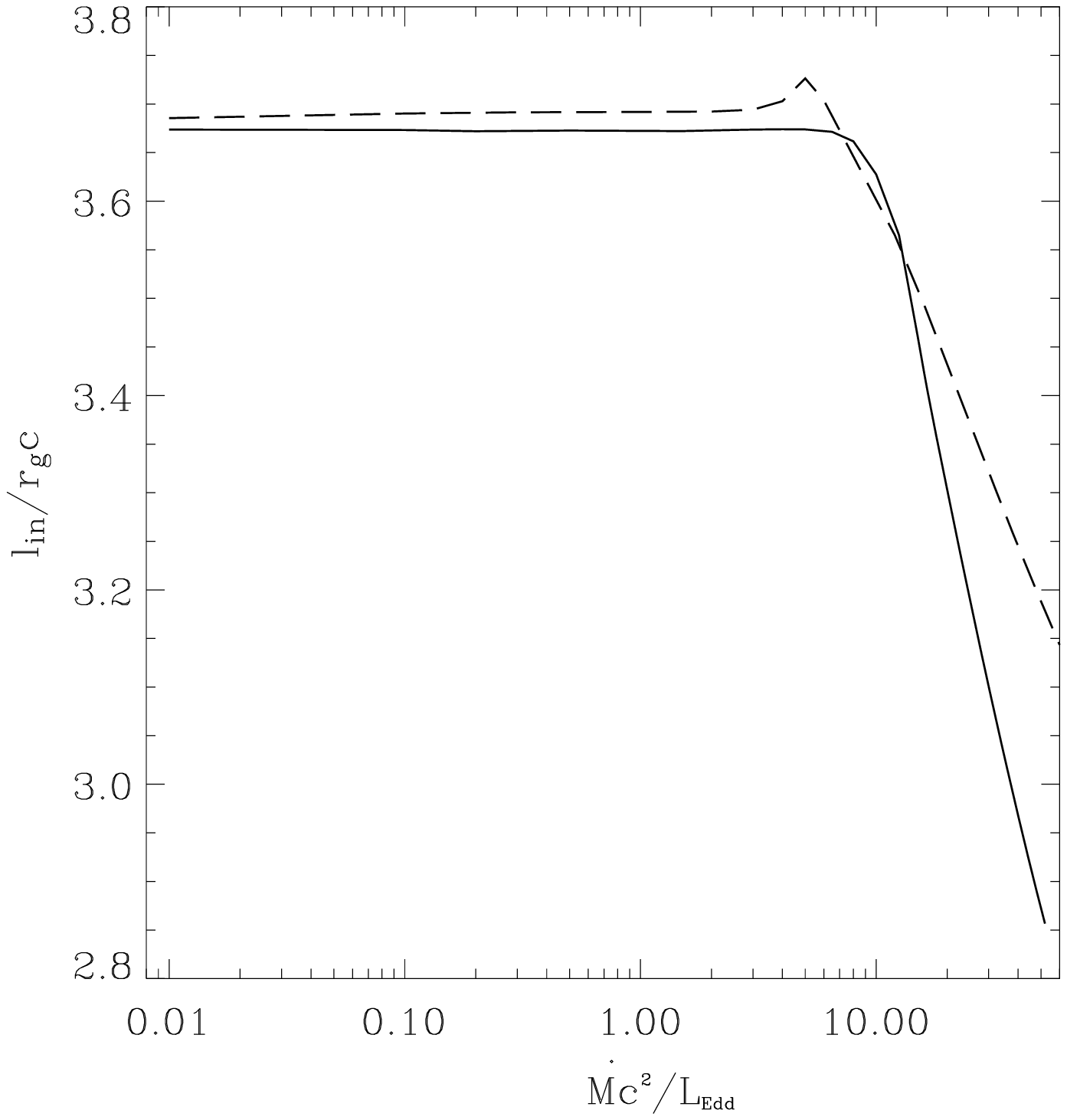}{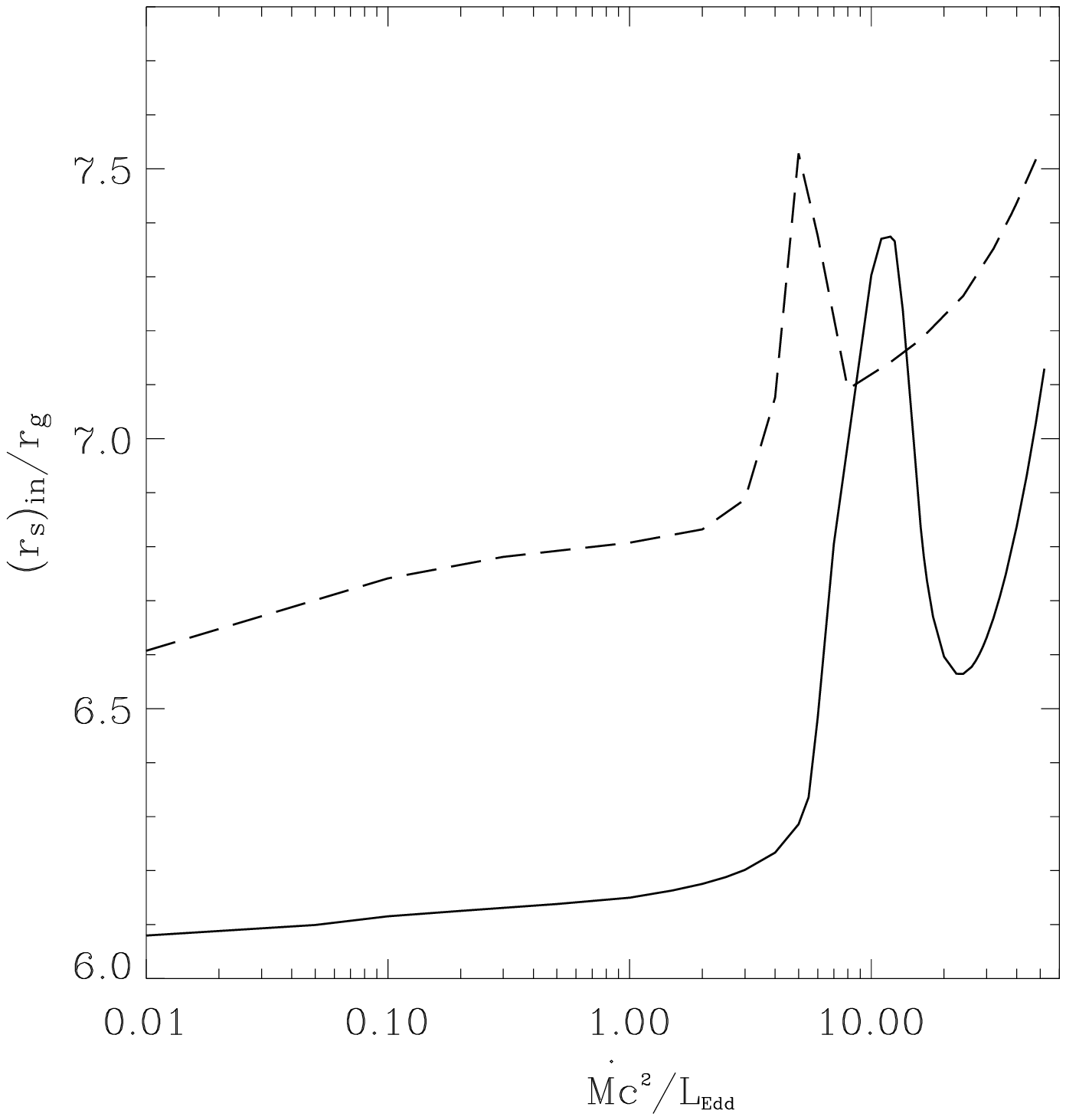}
\caption{The specific angular momentum
$\ell_{in}$ (left panel) and the position of the inner singular
points (right panel) as a function of the mass accretion rate
$\dot{M}$ for viscosity parameters $\alpha=0.5$ and
$M_{BH}=10M_{\odot}$. The solid lines represent models with the
optical depth transition and advection, whereas the dashed lines
correspond to advective optically thick ones.
\label{fig3}}
\end{figure}

\begin{figure}
\plotone{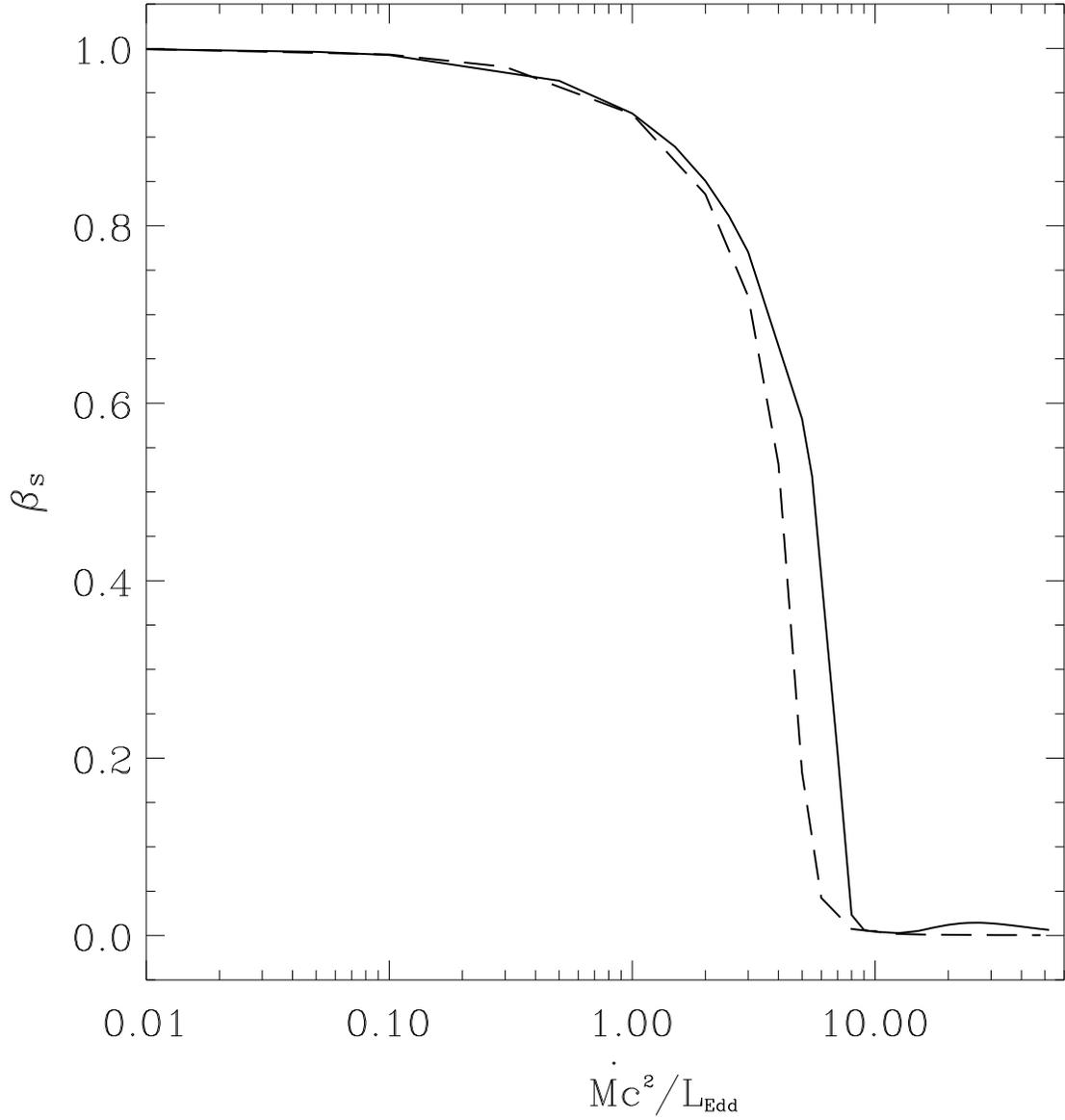}
\caption{Ratio of the gas pressure to the total
pressure at the inner singular point, $\beta_s$, as a function of
the mass accretion rate $\dot{M}$. Models with $\alpha=0.5$ and
$M_{BH}=10M_{\odot}$ are shown.
The styles of the curves are the same as in Fig.3.
\label{fig4}}
\end{figure}
%
%
\begin{figure}
\plotone{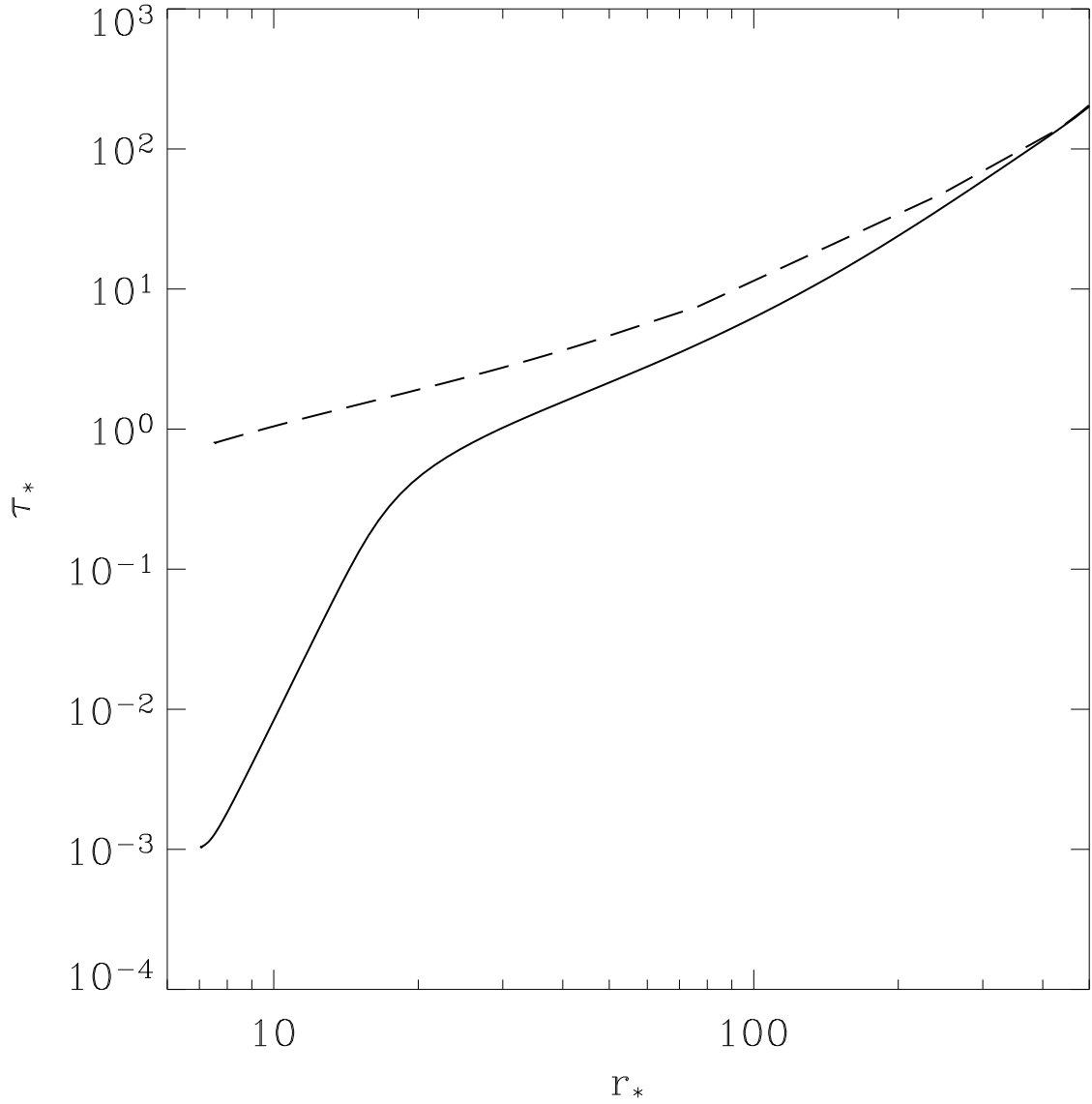}
\caption{The dependence of the
effective optical depth on the radius for
$M_{BH}=10M_{\odot}$, $\dot{m}=48$,
$\alpha=0.5$. Dashed line corresponds to the solution with
advection and without optical depth transition, solid line
corresponds to the solution with advection and optical depth
transition.
\label{fig5}}
\end{figure}
\begin{figure}
\plotone{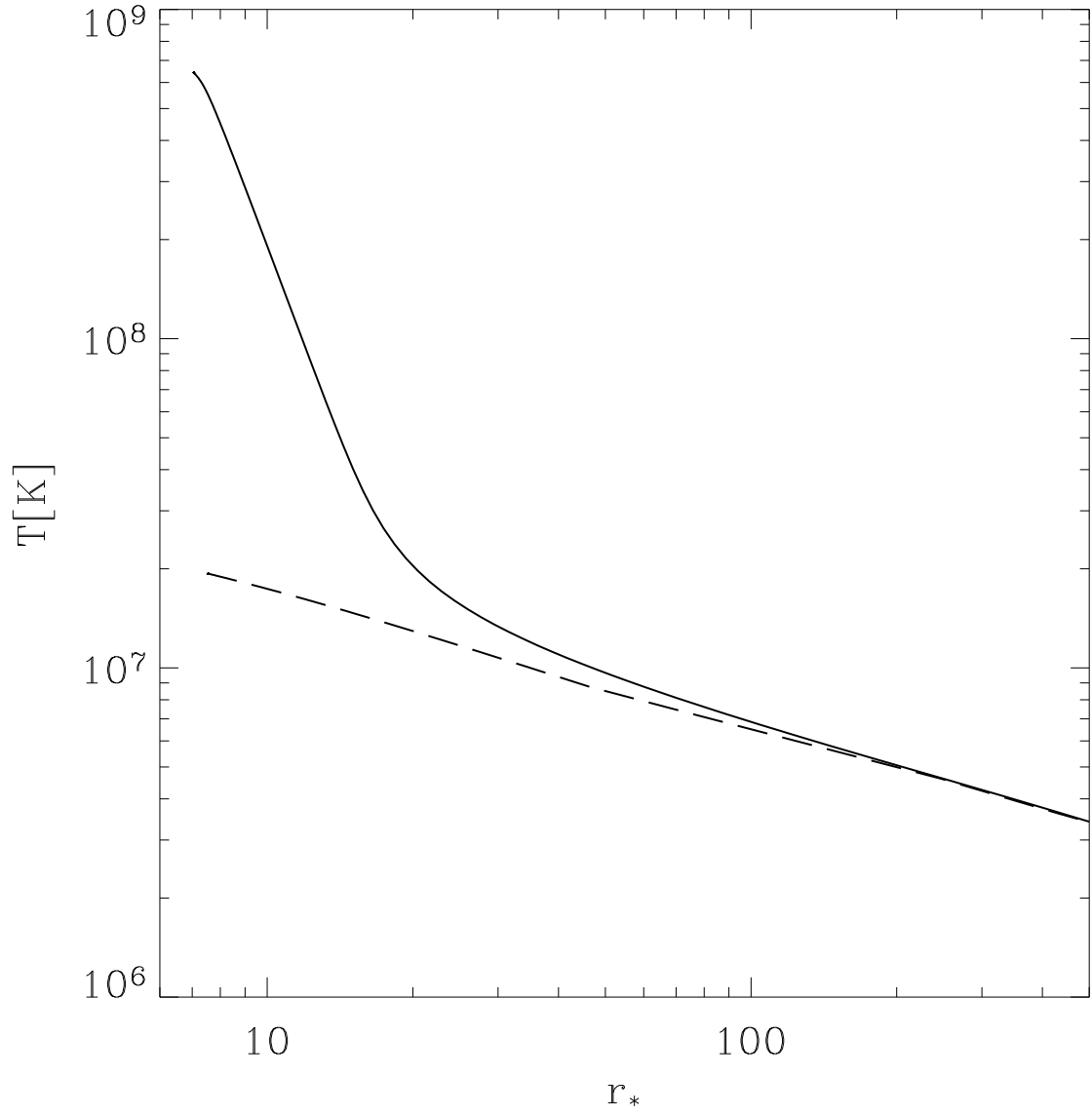}
\caption{The dependence of the temperature on the
radius for the same models as in Fig.5.
The styles of the curves are the same as in Fig.5.
\label{fig6}}
\end{figure}
\begin{figure}
\plotone{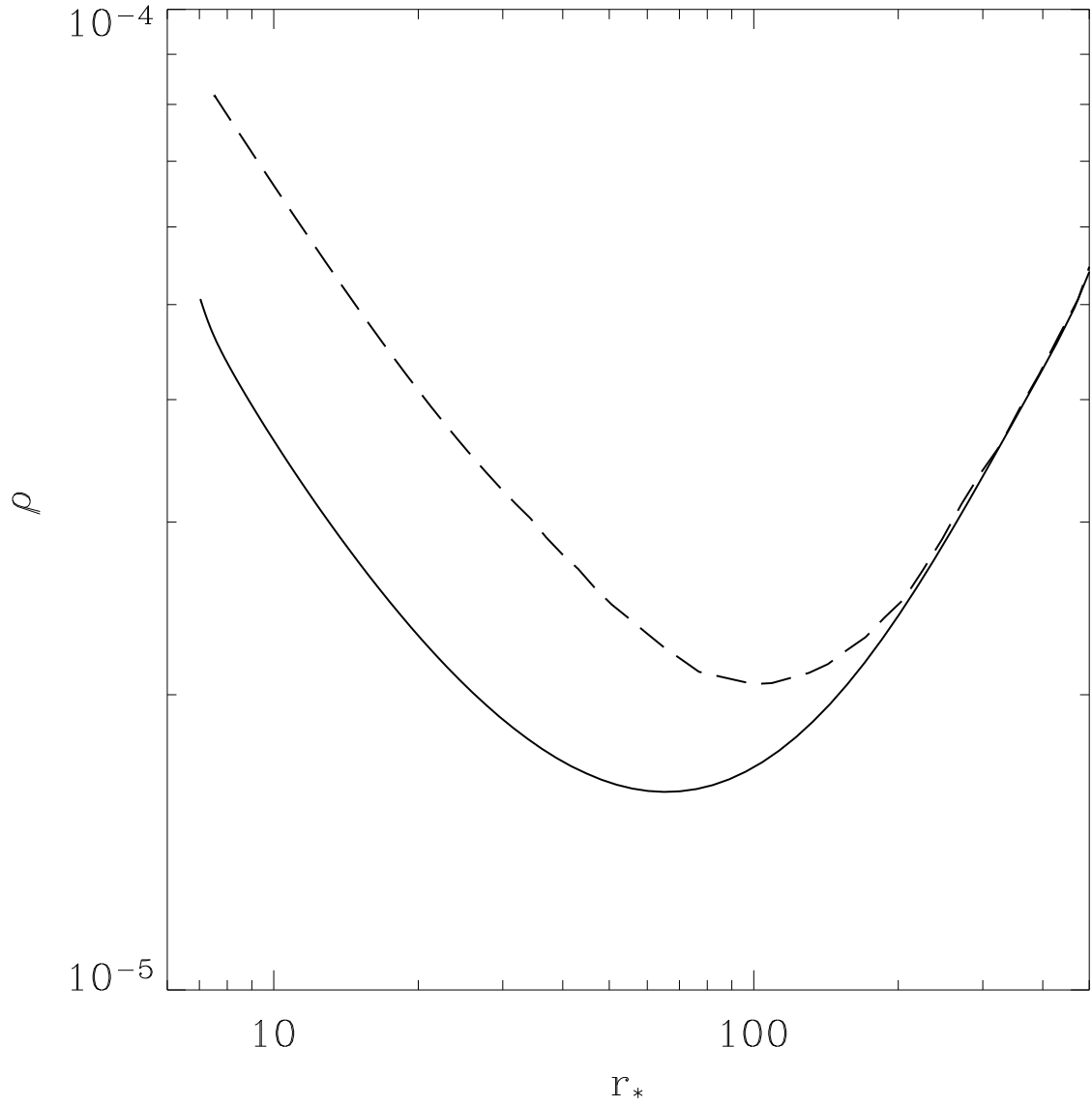}
\caption{The dependence of the density on the
radius for the same models as in Fig.5.
The styles of the curves are the same as in Fig.5.
\label{fig7}}
\end{figure}
\end{document}